\documentclass[10pt,twocolumn,letterpaper]{article}

\usepackage{titling} %
\usepackage{cvpr}
\usepackage{times}
\usepackage{epsfig}
\usepackage{graphicx}
\usepackage{amsmath}
\usepackage{amssymb}
\usepackage{bm}
\usepackage{subcaption}
\usepackage{booktabs,tabularx,tabulary}

\usepackage[square,numbers,sort,compress]{natbib}

\usepackage[pagebackref=true,breaklinks=true,colorlinks,bookmarks=false]{hyperref}

\cvprfinalcopy %

\def\cvprPaperID{1163} %

\ifcvprfinal\fi
\usepackage{overpic}
\usepackage{amssymb}
\usepackage{amsmath}
\usepackage{graphicx}
\usepackage{enumitem}
\usepackage{microtype} %

\usepackage{color}
\definecolor{turquoise}{cmyk}{0.65,0,0.1,0.3}
\definecolor{purple}{rgb}{0.65,0,0.65}
\definecolor{dark_green}{rgb}{0, 0.5, 0}
\definecolor{green}{rgb}{0, 1.0, 0}
\definecolor{orange}{rgb}{0.8, 0.6, 0.2}
\definecolor{red}{rgb}{0.8, 0.2, 0.2}
\definecolor{blueish}{rgb}{0.0, 0.7, 1}
\definecolor{light_gray}{rgb}{0.7, 0.7, .7}
\definecolor{pink}{rgb}{1, 0, 1}

\newcommand{\hide}[1]{{}} %

\newcommand{\Fig}[1]{\autoref{fig:#1}}
\newcommand{\Figure}[1]{\autoref{fig:#1}}
\newcommand{\Table}[1]{\autoref{tab:#1}}

\newcommand{\Sec}[1]{Section~\ref{sec:#1}}
\newcommand{\Section}[1]{Section~\ref{sec:#1}}

\usepackage{currfile}

\newcommand{\CIRCLE}[1]{\raisebox{.5pt}{\footnotesize \textcircled{\raisebox{-.6pt}{#1}}}}
\usepackage{amsmath}

\usepackage{lipsum}

\renewcommand{\paragraph}[1]{{\vspace{.25em}\noindent \textbf{#1.}}}

\newcommand{\volume}{\mathcal{V}}
\newcommand{\surface}{\mathcal{S}}
\newcommand{\texture}{\mathcal{T}}
\newcommand{\decodedsurface}{\mathcal{\hat{S}}}

\newcommand{\tsdf}{\Phi}
\newcommand{\threshold}{\tau}
\newcommand{\volumeWidth}{W}
\newcommand{\volumeHeight}{H}
\newcommand{\volumeDepth}{D}
\newcommand{\volumeDomain}{{\volumeWidth{}{\times}\volumeHeight{}{\times} \volumeDepth}}
\newcommand{\encoder}{\mathcal{E}}
\newcommand{\decoder}{\mathcal{D}}

\newcommand{\R}{\mathbb{R}}
\newcommand{\x}{\mathbf{x}} %
\newcommand{\block}{\bm{x}}
\newcommand{\blocksize}{{k \times k \times k}}

\newcommand{\code}{\mathbf{z}}

\newcommand{\quantizedcode}{\hat{\code}}

\newcommand{\sign}{\mathbf{s}}

\newcommand{\mortonize}{\mathbf{M}}
\newcommand{\chamfer}{\mathbf{C}}
\newcommand{\hausdorff}{\mathbf{H}}

\newcommand{\losssign}{R_{\sign}}
\newcommand{\lossrate}{R_{\quantizedcode}}
\newcommand{\lossdist}{D_{\hat{\block}}}
\newcommand{\ptheta}{\bm{\theta}}
\newcommand{\pphi}{\bm{\phi}}

\DeclareMathOperator*{\argmin}{arg\,min}

\newcommand{\supplementary}{{\color{dark_green}\textbf{supplementary material}}}
\newcommand{\video}{{\color{dark_green}\textbf{supplementary video}}}

\begin{document}

\title{Deep Implicit Volume Compression}

\author{Danhang Tang\thanks{indicates equal contribution.} \quad Saurabh Singh\footnotemark[1]
\quad Philip A.~Chou \quad Christian H{\"a}ne \quad 
Mingsong Dou \\ Sean Fanello \quad Jonathan Taylor \quad Philip Davidson \quad Onur G.~Guleryuz
\quad Yinda Zhang \\ Shahram Izadi \quad Andrea Tagliasacchi \quad Sofien Bouaziz \quad Cem Keskin
\\[.5em]
Google
}

\maketitle

\begin{abstract}
We describe a novel approach for compressing truncated signed distance fields (TSDF) stored in 3D voxel grids, and their corresponding textures. To compress the TSDF, our method relies on a block-based neural network architecture trained end-to-end, achieving state-of-the-art rate-distortion trade-off. To prevent topological errors, we losslessly compress the signs of the TSDF, which also upper bounds the reconstruction error by the voxel size. To compress the corresponding texture, we designed a fast block-based UV parameterization, generating coherent texture maps that can be effectively compressed using
existing video compression algorithms.
We demonstrate the performance of our algorithms on two 4D performance capture datasets, reducing bitrate by $66\%$ for the same distortion, or alternatively reducing the distortion by $50\%$ for the same bitrate, compared to the state-of-the-art.
\end{abstract}
\section{Introduction}
In recent years, volumetric implicit representations have been at the heart of many 3D and 4D reconstruction approaches~\cite{kinfu,dynfu,relightables,m2f}, enabling novel applications such as real time dense surface mapping in AR devices and free-viewpoint videos. 
While these representations exhibit numerous advantages, transmitting high quality 4D sequences is still a challenge due to their large memory footprints. 
Designing efficient compression algorithms for implicit representations is therefore of prime importance to enable the deployment of novel consumer-level applications such as VR/AR telepresence~\cite{holo}, and to facilitate the streaming of free-viewpoint videos~\cite{fvvbook}.

\begin{figure}
\centering
\includegraphics[trim={0cm 4cm 15.5cm 0cm},clip,width=1.\linewidth]{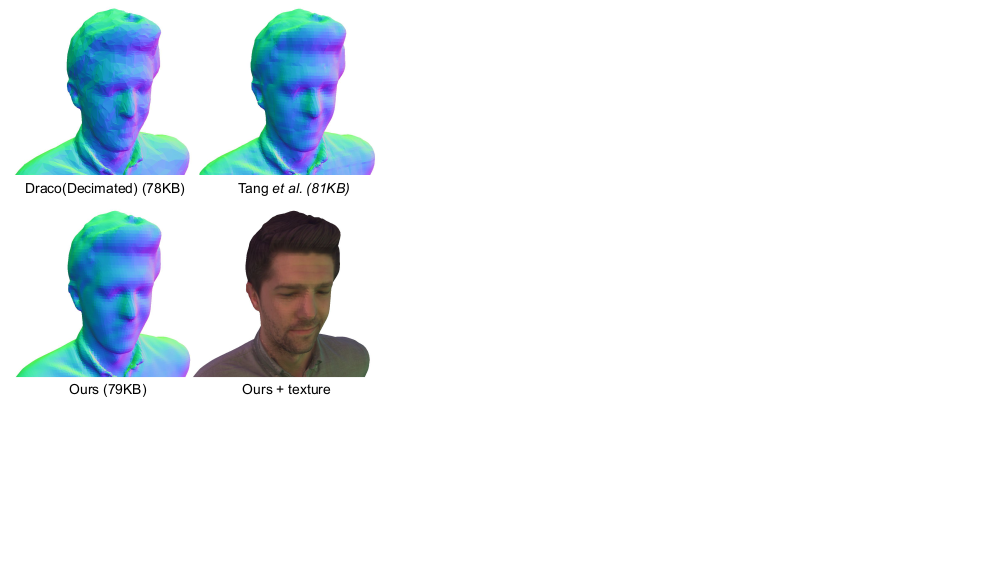}
\vspace{-2em}
\caption{When targeting a low bitrate, Draco~\cite{draco} requires decimation to have low-poly meshes as input, while \cite{compression} suffers from block artifacts. Our method has visibly lower distortion while maintaining similar bitrates. Raw meshes with flat shading are shown to reveal artifacts. }
\label{fig:teaser}
\vspace{-1em}
\end{figure}

In contrast to compressing a mesh, it was recently shown that truncated signed distance fields (TSDF) \cite{CurlessL:96} are highly suitable for efficient compression \cite{compression, KrivokucaKC:18} due to correlation in voxel values and their regular grid structure.
Voxel-based SDF representations have been used with great success for 3D shape learning using encoder-decoder architectures~\cite{conf/cvpr/SitzmannTHNWZ19,Wang:2017:OOC:3072959.3073608}.
This is in part due to their grid structure that can be naturally processed with 3D convolutions, allowing the use of convolutional neural networks (CNN) that have excelled in image processing tasks.
Based on these observations, we propose a novel block-based encoder-decoder neural architecture trained end-to-end, achieving bitrates that are $33\%$ of prior art~\cite{compression}. 
We compress and transmit the TSDF signs \textit{losslessly}; this does not only guarantee that the reconstruction error is upper bounded by the voxel size, but also that the reconstructed surface is \textit{homeomorphic}~{--}~even when lossy TDSF compression is used.
Furthermore, we propose using the conditional distribution of the signs given the encoded TSDF block to compress the signs losslessly, leading to significant gains in bitrates.
This also significantly reduces artifacts in the reconstructed geometry and textures.

Recent 3D and 4D reconstruction pipelines not only reconstruct accurate geometry, but also generate high quality texture maps, \eg $4096{\times}4096$ pixels, that need to be compressed and transmitted altogether with the geometry~\cite{relightables}.
To complement our TSDF compression algorithm, we developed a fast parametrization method based on block-based charting, which encourages spatio-temporal coherence without tracking.
Our approach allows efficient compression of textures using existing image-based techniques and \textit{removes} the need of compressing and streaming UV coordinates.

To summarize, we propose a novel block-based 3D  compression model with these contributions:
\begin{enumerate}[leftmargin=*,noitemsep]
\item the first deep 3D compression method that can train end-to-end with entropy encoding, yielding state-of-the-art performance;
\item lossless compression of the surface topology using the conditional distribution of the TSDF signs, and thereby bounding the reconstruction error by the size of a voxel;
\item a novel block-based texture parametrization that inherently encourages temporal consistency, without tracking or the necessity of UV coordinates compression.
\end{enumerate}

\section{Related works}
Compression of 3D/4D media (\eg, meshes, point clouds, volumes) is a fundamental problem for applications such as VR/AR, yet has received limited attention in the computer vision community.
In this section, we describe two main aspects of 3D compression: geometry and texture, 
as well as reviewing recent trends in learnable compression.

\begin{figure*}[t!]
\centering
\includegraphics[width=1.0\textwidth]{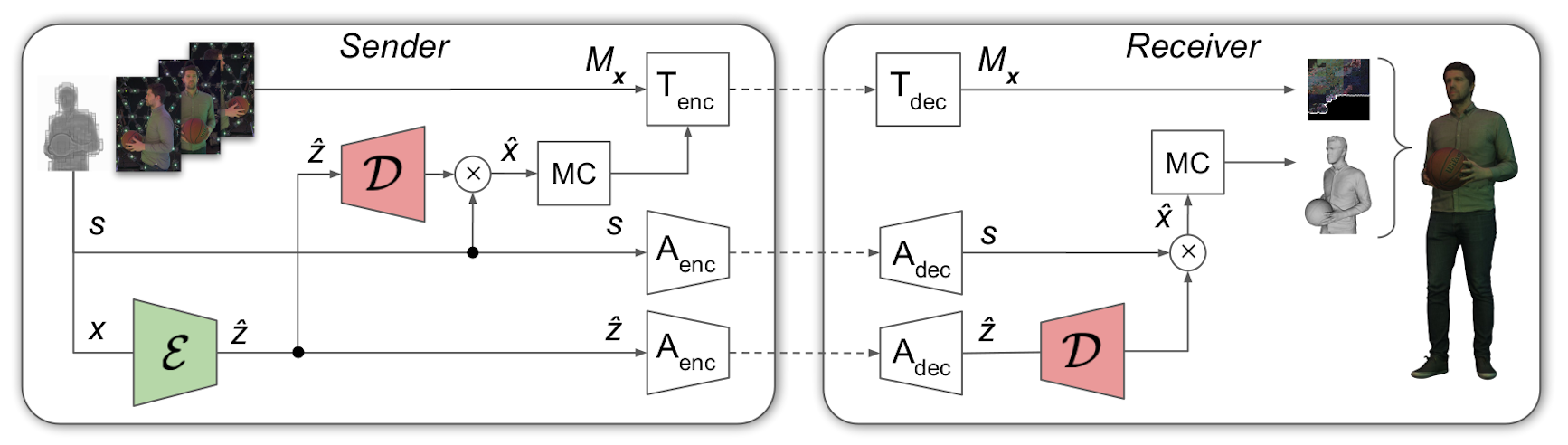}
\vspace{-2em}
\caption{\textbf{Compression pipeline --}
Given an input TSDF block $\block$ and its sign configuration $\sign {=} \operatorname{sign}(\block)$, an encoder transforms $\block$ into a quantized code $\quantizedcode {=} \lfloor \encoder(\block) \rceil$.
Then $\quantizedcode$ and $\sign$ are entropy coded and transmitted to the receiver ($A_{\text{enc}}$ and $A_{\text{dec}}$ blocks) using a prior learned distribution $p_{\quantizedcode}(\quantizedcode)$ and the conditional distribution $p_{\sign|\quantizedcode}(\sign|\quantizedcode)$ as estimated by the decoder, respectively.
The reconstructed block $\hat{\block}{=}\sign \odot |\decoder(\quantizedcode)|$ is used with marching cubes (MC in the figure) to extract the mesh, which is then used to generate the Morton packed chart $M_x$. $M_x$ is coded separately (with the $T_{\text{enc}}$ and $T_{\text{dec}}$ blocks).
}
\vspace{-1em}
\label{fig:pipeline}
\end{figure*}

\paragraph{Geometry compression}
Geometric surface representations can either be \textit{explicit} 
or  \textit{implicit}. 
While explicit representations are dominant in traditional computer graphics~\cite{pmp, fvv}, implicit representations have found widespread use in perception related tasks such as real-time volumetric capture~\cite{motion2fusion, kinfu, dynfu, dou2016fusion4d}.
\textit{Explicit} representations include meshes, point clouds, and parametric surfaces (NURBS). 
We refer the reader to the relevant surveys~\cite{AlliezG05,Peng_2005,maglo20153d} for compression of such representations.
Mesh compressors such as Draco~\cite{draco} use connectivity compression~\cite{Rossignac_1999,MamouZP09} followed by vertex prediction~\cite{gotsman-touma-gi98}.
An alternate strategy is to encode the mesh as geometry images \cite{GuGH02}, or geometry videos \cite{BricenoSMGH03} for temporally consistent meshes. 
Point clouds have been compressed by Sparse Voxel Octrees (SVOs)~\cite{JackinsT80,meagher_octtree}, first used for point cloud geometry compression in~\cite{schnabel_2006}.
SVOs have been extended to coding dynamic point clouds in~\cite{Kammerl2012} and implemented in the Point Cloud Library~(PCL)~\cite{Rusu_3dis}.
A version of this library became the anchor (\ie, reference proposal) for the MPEG Point Cloud Codec~(PCC)~\cite{MekuriaBC:17}.
The MPEG PCC standard is split into video-based PCC (V-PCC) and geometry-based PCC (G-PCC)~\cite{schwarz2018emerging}.
V-PCC uses geometry video, while G-PCC uses SVOs.
\textit{Implicit} representations include (truncated) signed distance fields (SDFs)~\cite{CurlessL:96} and occupancy/indicator functions~\cite{poisson}.
These have proved popular for 3D surface reconstruction~\cite{CurlessL:96, dynfu, DouTFFI15, dou2016fusion4d, m2f, compression, LoopCOC:16} and general 2D and 3D representation~\cite{Frisken:2000:ASD:344779.344899}.
Implicit functions have recently been employed for geometry compression~\cite{KrivokucaCK:18, canhelas, compression}, where the TSDF is encoded directly.

\paragraph{Texture compression}
In computer graphics, textures are images associated with meshes through UV maps.
These images can be encoded using standard image or video codecs~\cite{draco}.
For point clouds, color is associated with points as attributes.
Point cloud attributes can be coded via spectral methods~\cite{zhang_icip_2014,ThanouCF16,CohenTV16,QueirozC:17} or transform methods~\cite{QueirozC:16}.
Transform methods are used in MPEG G-PCC~\cite{schwarz2018emerging}, and, similarly to TSDFs, have volumetric interpretation~\cite{ChouKK:18}.
Another approach is to transmit the texture as ordinary video from each camera, and use projective texturing at the receiver~\cite{compression}.
However, the bitrate increases linearly with the number of cameras, and projective texturing can create artifacts when the underlying geometry is compressed.
Employing a UV parametrization to store textures is not trivial, as enforcing spatial and temporal consistency can be computationally intensive.
On one end of the spectrum, Motion2Fusion~\cite{m2f} sacrifices the spatial coherence typically desired by simply mapping each triangle to an arbitrary position of the atlas, hence sacrificing compression rate for efficiency.
On the other extreme,~\cite{prada17, relightables} take a step further by tracking features over time to generate a temporally consistent mesh connectivity and UV parametrization, therefore can be compressed with modern video codecs. This process is however expensive and cannot be applied to real-time applications.

\paragraph{Learnable compression strategies}
Learnable compression strategies have a long history.
Here we focus specifically on neural compression.
The use of neural networks for image compression can be traced back to 1980s with auto-encoder models using uniform~\cite{munro1989image} or vector~\cite{luttrell1988image} quantization. 
However, these approaches were akin to non-linear dimensionality reduction methods and do not learn an entropy model explicitly.
More recently~\citet{toderici15rnn} used a recurrent LSTM based architecture to train multi-rate progressive coding models.
However, they learned an explicit entropy model as a separate post processing step after the training of recurrent auto-encoding model.
\citet{balle2017end} proposed an end-to-end optimized image compression model that jointly optimizes the rate-distortion trade-off.
This was extended by placing a hierarchical hyperprior on the latent representations to significantly improve the image compression performance~\cite{hyperprior}.
While there has been significant application of deep learning on 3D/4D representations, \eg~\cite{conf/cvpr/WuSKYZTX15, qi2016pointnet, Wang:2017:OOC:3072959.3073608,conf/cvpr/LiaoDG18, conf/cvpr/SitzmannTHNWZ19, conf/cvpr/ParkFSNL19},
application of deep learning to 3D/4D \textit{compression} has been scant. 
However, very recent works closely related to ours have used rate-distortion optimized auto-encoders similar to~\cite{hyperprior} to perform 3D geometry compression end-to-end:
\citet{yan2019deep} used a PointNet-like encoder combined with a fully-connected decoder, trained to minimize directly the Chamfer distance subject to a rate constraint, on the entire point cloud.
\citet{quach2019learning} performs block-based coding to obtain higher quality on the %
MVUB dataset~\cite{LoopQOC:17}.
Their network predicts voxel occupancy using a \textit{focal loss}, which is similar to a weighted binary cross entropy.
In the most complete and performant work until now, \citet{wang2019learned} also uses block-based coding and predicted voxel occupancy, with a weighted binary cross entropy.
They reported a 60\% bitrate reduction compared to MPEG G-PCC on the high resolution 8iVFB dataset~\cite{dEonHMC:17} hosted by MPEG, though they report only approximate equivalence with state-of-the-art MPEG V-PCC.  

In contrast, we use block-based coding on even higher resolution datasets, and report bitrates that are at least three times better than MPEG V-PCC, by compressing the TSDF directly rather than occupancy, yielding sub-voxel precision.

\section{Background}
\label{sec:outline}
Our goal is to compress an input sequence of TSDF volumes $\volume{=}\{\volume_t\}_1^T$ encoding the geometry of the surface, and their corresponding texture atlases $\texture{=}\{\texture_t\}_1^T$, which are both extracted from a multi-view RGBD sequence~\cite{compression, relightables}.
Since geometry and texture are quite different, we compress them separately. The two data streams are then fused by the receiver before rendering. To compress the geometry data $\volume$, inspired by the recent advances in learned compression methods, we propose an end-to-end trained compression pipeline taking volumetric blocks as input; see~\Section{geometry}. Accordingly we also design a block-based UV parametrization algorithm for texture $\texture$; see~\Section{compression}.
For those unfamiliar with the topic and notation, we overview fundamentals of compression in the \supplementary{}.

\section{Geometry compression}
\label{sec:geometry}

There are two primary challenges in end-to-end learning of compression, both of which arise from the non-differentiability of intermediate steps:
\CIRCLE{1} compression is non-differentiable due to the quantization  necessary for compression;
\CIRCLE{2} surface reconstruction from TSDF values is typically non-differentiable in popular methods such as Marching Cubes~\cite{marchingcubes}.
To tackle~\CIRCLE{1}, we draw inspiration from the recent advances in learned image compression~\cite{hyperprior, balle2017end}.
To tackle~\CIRCLE{2}, we make the observation that Marching Cubes algorithm is differentiable with \emph{known topology}.

\paragraph{Computational feasibility of training}
The dense TSDF volume data $\volume{=}\{\volume_t\}_{t=1}^T$ for an entire sequence is very high dimensional. 
For example, a sequence from the dataset used in ~\citet{compression} has 500 frames, with each frame containing $240 \times 240 \times 400$ voxels.
The high dimensionality of data makes it computationally infeasible to compress the entire sequence jointly.
Therefore, following~\citet{compression}, we process each frame independently in a block based manner.
From the TSDF volume~$\volume$, we extract all non-overlapping \textit{blocks} $\{\block_m\}_{1}^{M}$ of size $\blocksize$ that contain a zero crossing. 
We refer to these blocks as \emph{occupied blocks}, and compress them independently.

\subsection{Inference}
\label{sec:inference}
The compression pipeline is illustrated in~\Figure{pipeline}.
Given a block $\block$ to be transmitted, the sender first computes the lossily quantized latent representation $\quantizedcode{=}\lfloor \encoder(\block; \ptheta_e) \rceil$ using the learned encoder $\encoder$ with parameters $\ptheta_e$. 
Next, the sender uses $\quantizedcode$ to compute the conditional probability distribution over the TSDF signs as $p_{\sign|\quantizedcode}(\sign | \quantizedcode; \ptheta_s)$, where $\sign$ is the ground truth sign configuration of the block, and $\ptheta_s$ are the learnable parameters of the distribution.
The sender then uses an entropy coder to compute the bitstreams $\quantizedcode_{\text{bits}}$ and $\sign_{\text{bits}}$ by losslessly coding the latent code $\hat{\code}$ and signs $\sign$ using the distributions $p_{\quantizedcode}(\quantizedcode; \pphi)$ and $p_{\sign|\quantizedcode}(\sign|\quantizedcode; \ptheta_s)$ respectively. Here $p_{\quantizedcode}(\quantizedcode; \pphi)$ is a learned prior distribution over $\quantizedcode$ parameterized by $\pphi$.
Note that while the prior distribution $p_{\quantizedcode}$ is part of the model and known a priori both to the sender and the receiver, the conditional distribution $p_{\sign|\quantizedcode}$ needs to be computed by both.
$\quantizedcode_{\text{bits}}$ and $\sign_{\text{bits}}$ are then transmitted to the receiver, which first recovers $\quantizedcode$ using entropy decoding with the shared prior $p_{\quantizedcode}$.  The receiver then re-computes $p_{\sign|\quantizedcode}$ in order to recover the losslessly coded ground truth signs $\sign$. 
Finally, the receiver recovers the lossy TSDF values by using the learned decoder~$\decoder$ in conjunction with the ground truth signs $\sign$ as $\hat{\block} = \sign \odot |\decoder(\quantizedcode; \ptheta_d)|$, where $\odot$ is the element--wise product operator, $|\cdot|$ the element--wise absolute value operator, and $\ptheta_d$ the parameters of the decoder.

\begin{figure}
\begin{center}
    \includegraphics[trim={0cm 6cm 0cm 0cm},clip,width=1.\linewidth]{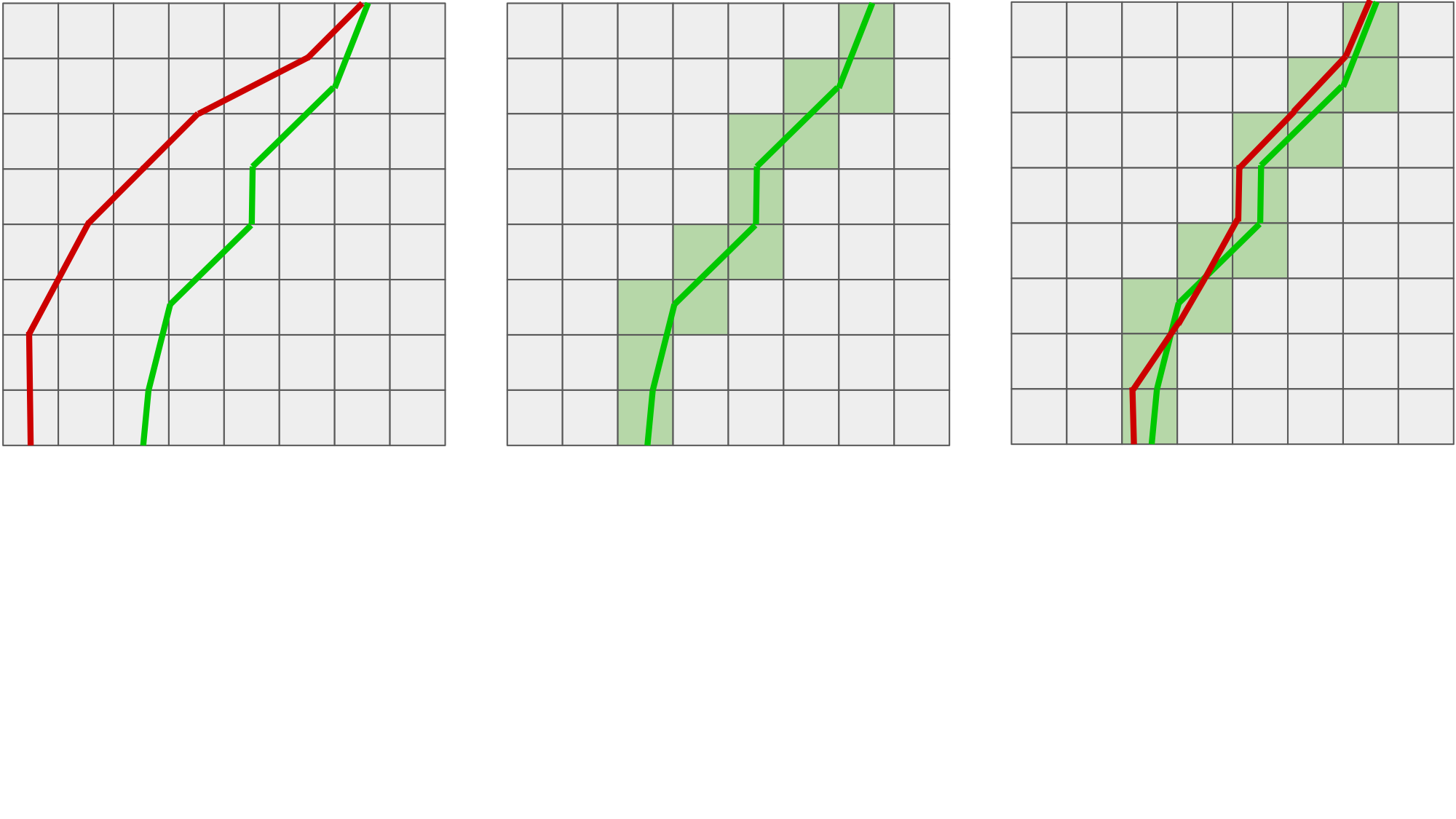}
    \vspace{-3em}
\end{center}
\caption{\textbf{Topology mask in inference}: 
We illustrate a 2D slice from a block, where each cell represents a voxel. (left) Without masking, the reconstructed surface (red) deviates from the ground truth (green) because of compression error. (mid) Losslessly compressed signs will give us ground truth occupancy/topology during inference. (right) Therefore, the average reconstructed error due to lossy magnitude compression is bounded by the size of a voxel ($5$mm).
}
\label{fig:gt_signs}
\vspace{-1em}
\end{figure}

To stitch the volume together, the block indices are transmitted to the client as well.
Similar to~\cite{compression}, the blocks are sorted in an ascending manner, and delta encoding is used to convert the vector of indices to a representation that is entropy encoder friendly. 
Once the TSDF volume is reconstructed, a triangular mesh can be extracted via marching cubes.
Note that for the marching cube algorithm, the polygon configurations are fully determined by the signs.
As we transmit the signs losslessly, it is \textit{guaranteed} that the mesh extracted from the decoded TSDF $\hat{\block}$  will have the same topology as the  mesh extracted from the uncompressed TSDF $\block$.
It follows that the only possible reconstruction errors will be at the vertices that lie on the edges of the voxels.
Therefore, the maximum reconstruction error is bounded by the edge length, \ie~the voxel size, as shown in~\Figure{gt_signs}.

\subsection{Training}
\label{sec:training}
We learn the parameters $\Theta{=}\{\ptheta_e, \ptheta_s, \ptheta_d, \pphi\}$ of our compression model by minimizing the following objective
\begin{align}
\argmin_{\Theta}
\:
\underbrace{\lossdist(\block, \hat{\block}; \ptheta_e, \ptheta_d)}_{\text{distortion}} +
\lambda [\underbrace{\lossrate(\quantizedcode; \pphi)}_{\text{latents bitrate}} +
\underbrace{\losssign(\sign; \ptheta_s)}_{\text{signs bitrate}}]
\label{eq:rate_dist_ours}
\end{align}

\paragraph{Distortion $\bm{\lossdist(\block, \hat{\block}; \ptheta_e, \ptheta_d)}$}
We minimize the reconstruction error between the ground truth and the predicted TSDF values.
However, directly computing the squared difference~$\|\hat{\block}~{-}~\block\|_2^2$ wastes model complexity on learning to precisely reconstruct values of TSDF voxels that are far away from the surface.
In order to focus the network on the important voxels (\ie the ones with a neighboring voxel of opposing sign), we use the ground truth signs.
For each dimension, we create a mask of important voxels, namely $m_x$, $m_y$ and $m_z$.
Voxels that have more than one neighbor with opposite signs appear in multiple masks, further increasing their weights.
We then use these masks to calculate the squared differences for important voxels only
$
 \lossdist = \tfrac{1}{B} \sum_{n=1}^B \sum_{d \in {x, y, z}} \|m_d \cdot (\hat{\block}_n - \block_n)\|_2^2,
$
for $B$ blocks.

\paragraph{Rate of latents $\bm{\lossrate(\quantizedcode; \pphi)}$}
A second loss term we employ is $\lossrate$, which is designed to reduce the bitrate of the compressed codes.
This loss is essentially a differentiable estimate of the non-differentiable Shannon entropy of the quantized codes $\quantizedcode$;~see~\cite{balle2017end} for additional details.

\paragraph{Rate of losslessly compressed signs $\bm{\losssign(\sign; \ptheta_s)}$}
Since~$\sign$ contains only discrete values $\{-1,+1\}$, it can be compressed losslessly using entropy coding.
As mentioned above, we use the conditional probability distribution~$p_{\sign|\quantizedcode}(\sign | \quantizedcode)$ instead of the prior distribution~$p_{\sign}(\sign)$.
Note that the conditional distribution should have a much lower entropy than the priors, since $\sign$ is dependent on the $\quantizedcode$ by design.
This allows us to compress the signs far more efficiently.

To make this dependency explicit, we add an extra head to the decoder, such that $p_{\sign}(\sign | \quantizedcode) {=} \decoder_s(\quantizedcode)$, and~$\hat{\block} {=} \sign\odot|\decoder_b(\quantizedcode)|$.
The sign rate loss $\losssign$ is then the cross entropy between the ground truth signs $\sign$, 
with $-1$ remapped to $0$, and their conditional predictions $p_{\sign}(\sign | \quantizedcode)$.
Minimizing $\losssign$ has the effect of training the network to make better sign predictions, while also minimizing the bitrate of the compressed signs.

\paragraph{Encoder and Decoder architectures}
Our proposed compression technique is agnostic to the choice of the individual architectures for the encoder and decoder.
In this work, we targeted a scenario requiring a maximum model size of roughly 2MB, which makes the network suitable for mobile deployment.
To limit the number of trainable parameters, we used convolutional networks, where both the encoder and the decoder consist of a series of 3D convolutions and transposed convolutions.
More details about the specific architectures can be found in the \supplementary{}.

\begin{figure*}[t!]
\begin{subfigure}[b]{0.66\textwidth}
\begin{center}
   \includegraphics[trim={1cm 3.5cm 3.5cm 1.5cm},clip, width=1.0\linewidth]{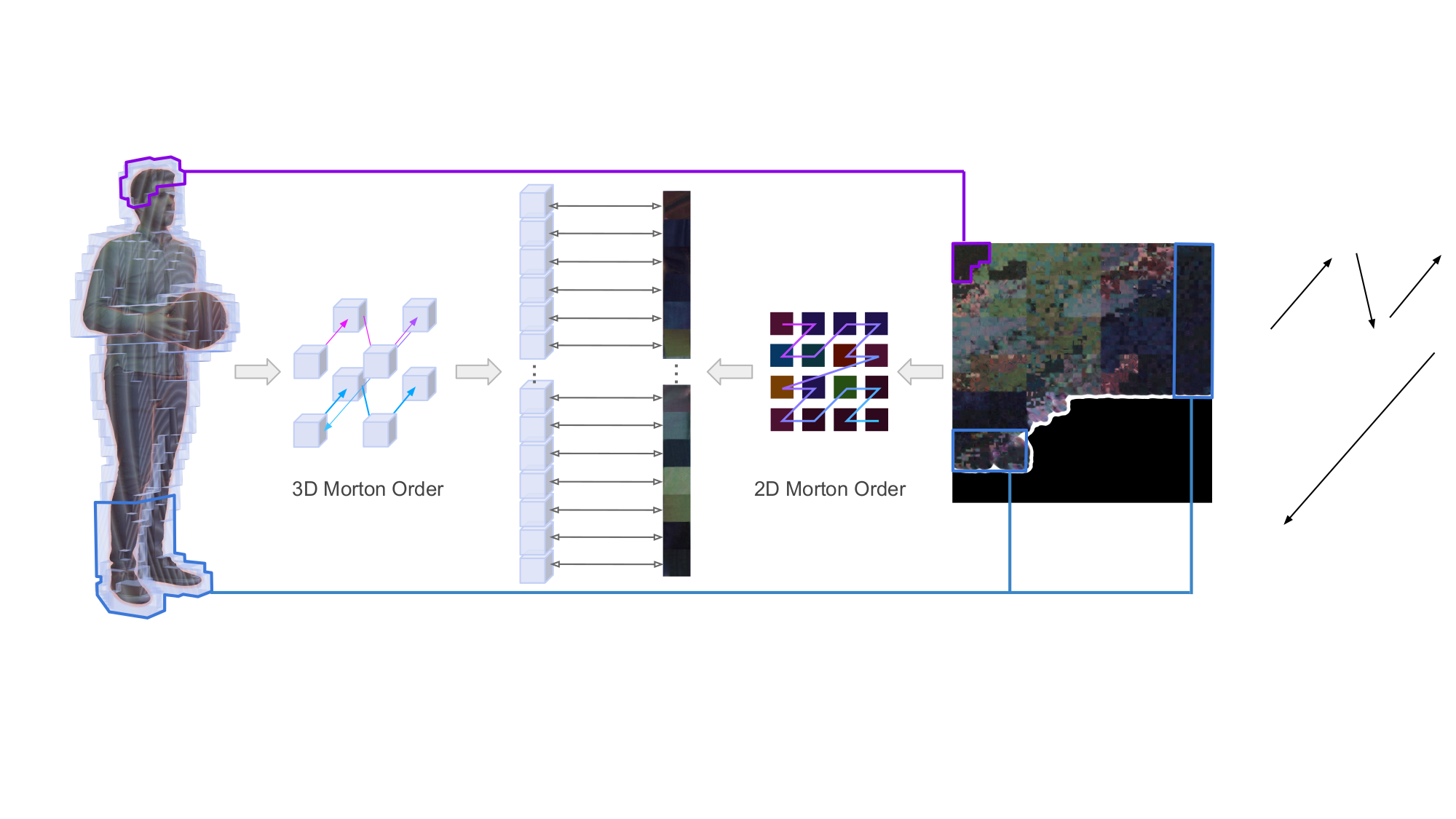}
\end{center}
\end{subfigure}
\begin{subfigure}[b]{0.33\textwidth}
\begin{center}
   \includegraphics[trim={0cm 0.7cm 8cm 0.7cm},clip, width=1.0\linewidth]{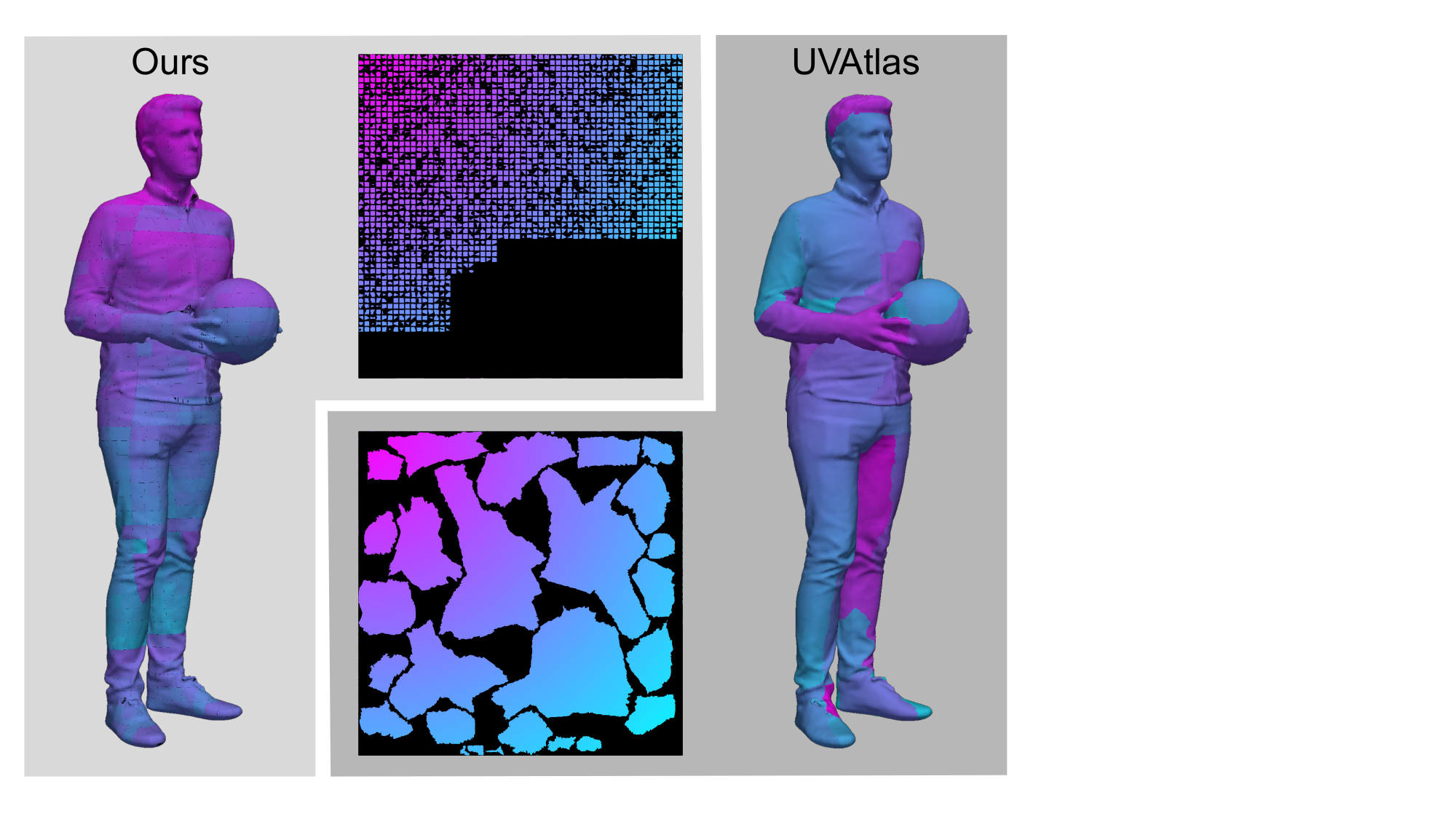}
\end{center}
\end{subfigure}
\vspace{-.5em}
\caption{\textbf{Texture packing --}
\label{fig:morton_packing}
(left) 3D blocks and 2D patches are ordered and matched by their Morton codes respectively.
This process unwraps the 3D volume to the texture atlas.
(right) The UVAtlas~\cite{zhou2004iso} only ensures local spatial coherence \textit{within} each chart, whilst our method encourages global spatial coherence.
Refer to the \video{} for a comparison on temporal coherence.
}
\vspace{-1em}
\end{figure*}

\section{Texture compression}
\label{sec:compression}
We propose a novel efficient and \textit{tracking-free} UV parametrization method to be seamlessly \textit{combined} with our block-level geometry compression; see \Figure{pipeline}. As our parametrization process is deterministic, UV coordinates can be inferred on the receiver side, thus removing the need for compression and transmission of the UV coordinates.

\paragraph{Block-level charting}
Traditional UV mapping either partitions the surface into a few large charts~\cite{zhou2004iso}, or generates one chart per triangle  
to avoid UV parametrization as in PTEX~\cite{burley2008ptex}. In our case, since the volume has already been divided into fixed-size blocks during geometry compression, it is natural to explore block-level parametrization.
To accommodate compression error, the compressed signal is decompressed on the sender side, such that both the sender and receiver have access to identical reconstructed volumes; see \Figure{pipeline} (left).
Triangles of each occupied block are then extracted and grouped by their normals.
Most blocks have only one group, while blocks in more complex areas (\eg fingers) may have more. 
The vertices of the triangles in each group are then mapped to UV space as follows:
\CIRCLE{1} the average normal in the group is used to determine a tangent space, onto which the vertices in the group are projected; \CIRCLE{2} the projections are rotated until they fit into an axis-aligned rectangle with minimum area, using rotating calipers \cite{Toussaint83solvinggeometric}.
This results in deterministic UV coordinates for each vertex in the group relative to a bounding box for the vertex projections; \CIRCLE{3} the bounding boxes for the groups in a block are then sorted by size and packed into a chart using a quadtree-like algorithm. There is exactly one 2D chart for each occupied 3D block.
After this packing, the UV coordinates for the vertices in the block are offset to be relative to the chart.
These charts are then packed into an atlas, where the UV coordinates for the vertices are again offset to be relative to the atlas, \ie to be a global UV mapping.
After UV parametrization, color information can be obtained from either per-vertex color in the geometry, previously generated atlas or even raw RGB captures.
Our method is \textit{agnostic} to this process.

\paragraph{Morton packing}
In order to optimize compression, the block-level charts need to be packed into an atlas in a way that maximizes \textit{spatio-temporal} coherence.
This is non-trivial, as in our sparse volume data structure the amount and positions of blocks can vary from frame to frame.
Assuming the movement of the subject is smooth, preserving the 3D spatial structure among blocks during packing is expected to preserve spatio-temporal coherence.
To achieve this effect we propose a Morton packing strategy.
Morton ordering \cite{Morton66} (also called Z-order curve) has been widely used in 3D graphics to create spatial representations~\cite{lauterbach2009fast}.
As our {\em blocks} are on a 3D regular grid, each occupied block can be indexed by a triple of integers $(x, y, z){\in}\mathbb{Z}^{3}$.
Each integer has a binary representation, \eg $x_{B-1}\cdots x_0$, where $x{=}\sum_{b=0}^{B-1}x_b2^b$.  The 3D Morton code for $(x,y,z)$ is defined as the integer $\mortonize_3(x,y,z){=}\sum_{b=0}^{B-1}(4y_b+2x_b+z_b)2^{3b}$ whose binary representation consists of the interleaved bits $y_{B-1}x_{B-1}z_{B-1}\cdots y_0x_0z_0$.  Likewise, as our {\em charts} are on a 2D regular grid, each chart can be indexed by a pair of integers $(u,v){\in}\mathbb{Z}^2$, whose 2D Morton code is the integer $\mortonize_2(u,v){=}\sum_{b=0}^{B-1}(2u_b+v_b)2^{2b}$ whose binary representation is $u_{B-1}v_{B-1}\cdots u_0v_0$.  These functions are invertible simply by demultiplexing the bits.  We map the chart for an occupied block at volumetric position $(x,y,z)$ to atlas position $(u,v){=}\mortonize_2^{-1}(rank(\mortonize_3(x,y,z)))$,
where $rank$ is the rank of the 3D Morton code in the list of 3D Morton codes,
as illustrated in \Figure{morton_packing}~(left).
Note that we \textit{choose} to prioritize $y$ over $x$ and $z$ when interleaving their bits into the 3D Morton code,
as $y$ is the vertical direction in our coordinate system, to accommodate typically standing human figures.
Hence, as long as blocks move smoothly in 3D space, corresponding patches are likely to move smoothly in the atlas, leading to an approximate spatio-temporal coherence, and therefore better (video) texture compression efficacy.

\vspace{-.5em}
\section{Evaluation}
To assess our method, we rely on the dataset captured by \citet{compression}, which consists of six ${\sim}500$ frames long RGBD multi-view sequences of different subjects at~$30$Hz. 
We use three of them for training and the others for evaluation.
We also employ ``The Relightables'' dataset by~\citet{relightables}, which contains higher quality geometry and higher resolution texture maps -- three ${\sim}600$-frame sequences.
To demonstrate the \textit{generalization} of learning-based methods, we only train on the dataset~\citet{compression}, and test on both ~\citet{compression} and \citet{relightables}.

\subsection{Geometry compression}
\label{sec:geometry_experiments}
We evaluate geometry compression using two different metrics: the Hausdorff metric ($\hausdorff$)~\cite{metro} measures the ($\max$) \textit{worst-case} reconstruction error via:
\begin{equation}
    \hausdorff(\surface, \decodedsurface) =
    \max \left(\max_{x \in \surface_v} d(x, \decodedsurface), \max_{y \in \decodedsurface_v} d(y, \surface) \right),
\end{equation}
where $\surface_v$ and $\decodedsurface_v$ are the set of points on the ground truth and decoded surface respectively. $d(\x, \surface)$ is the shortest Euclidean distance from a point $\x{\in}\R^3$ to the surface $\surface$. Another metric is the symmetric Chamfer distance ($\chamfer$):
\begin{equation}
\chamfer(\surface, \decodedsurface) =
\tfrac{1}{2|\surface_v|} \sum_{x \in \surface_v} d(x, \decodedsurface) +
\tfrac{1}{2|\decodedsurface_v|} \sum_{y \in \decodedsurface_v} d(y, \surface).
\end{equation}
For each metric, we compute a final score averaging all volumes, which we refer to as \textit{Average Hausdorff Distance} and \textit{Average Chamfer Distance} respectively.

\begin{table}[t]
\centering
\begin{tabularx}{0.475\textwidth}{Xccc}
\toprule
 & \textbf{Raw data} & \textbf{Na\"ive} & \textbf{Ours} \\
\midrule
\textbf{Avg. Size / Volume} & 155.1KB & 139.8KB & \textbf{2.9KB} \\
\bottomrule   
\end{tabularx}
\vspace{-.5em}
\caption{\textbf{Lossless sign compression}: Our data-driven probability model, combined with an arithmetic coder, can improve the compression rate by \textbf{48$\times$} comparing to a na\"ive probabilty model based on statistics of signs in the dataset.}
\vspace{-1em}
\label{tab:sign_compression}
\end{table}

\paragraph{Signs} 
We showcase the benefit of our data dependent probability model on rate in~\Table{sign_compression}. Raw sign data, though being binary, has an average size of $154.1$KB per volume. With na\"ively computed probability of signs being positive over the dataset, an arithmetic coder can slightly improve the rate to $139.8$ KB. This is because there are more positive TSDF values than negative in the dataset. With our learned, data dependent probability model, the arithmetic coder can drastically compress the signs down to $2.9$ KB per volume.

\begin{figure}
\begin{subfigure}{0.49\linewidth}
\begin{center}
\includegraphics[width=1.0\linewidth]{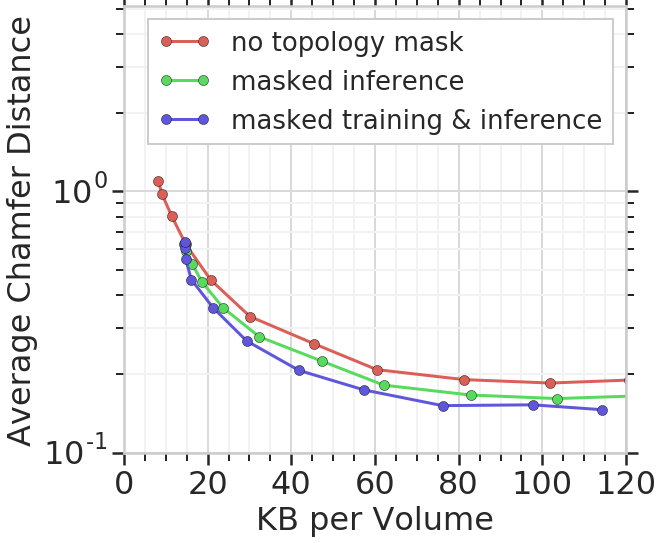}
\vspace{-1em}
\end{center}
\end{subfigure}
\begin{subfigure}{0.49\linewidth}
\begin{center}
\includegraphics[width=1.0\linewidth]{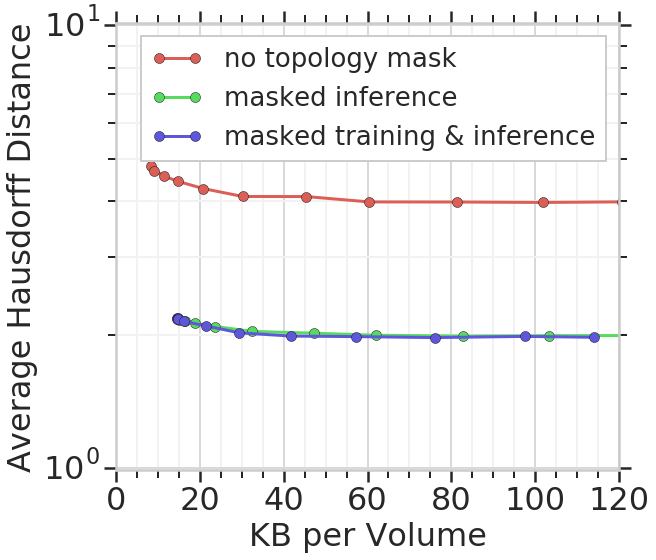}
\vspace{-1em}
\end{center}
\end{subfigure}
\vspace{-1em}
\caption{\textbf{Topology Mask}: 
When topology masking is applied during inference, an upper bound of error is guaranteed. Moreover, when also applied as a training loss, topology mask yields better rate-distortion. The difference is more obvious with the Hausdorff distance, which measures the worst case error.
}
\vspace{-1em}
\label{fig:exp_topology_mask}
\end{figure}

\paragraph{Topology Masking}
To demonstrate the impact of utilizing ground truth sign/topology, we construct a baseline with a standard rate-distortion loss. %
Specifically, the distortion term is simplified as
$
 \lossdist {=} \tfrac{1}{B} \sum_{n=1}^B  \| \hat{\block}_n - \block_n\|_2^2.
$
This baseline is shown as \texttt{no topology mask} in \Fig{exp_topology_mask}. Without the error bound, %
its distortion is much higher than other baselines. %
The second baseline, in addition to using the same distortion term, losslessly compresses and streams the signs during inference, as described in \Sec{geometry}.
Despite the increased rate due to losslessly compressed signs, this baseline still achieves better rate-distortion trade-off.
Finally, using topology masking in both training and inference yields the best rate-distortion performance.

\begin{figure}
\begin{subfigure}{0.49\linewidth}
\begin{center}
\includegraphics[width=1.0\linewidth]{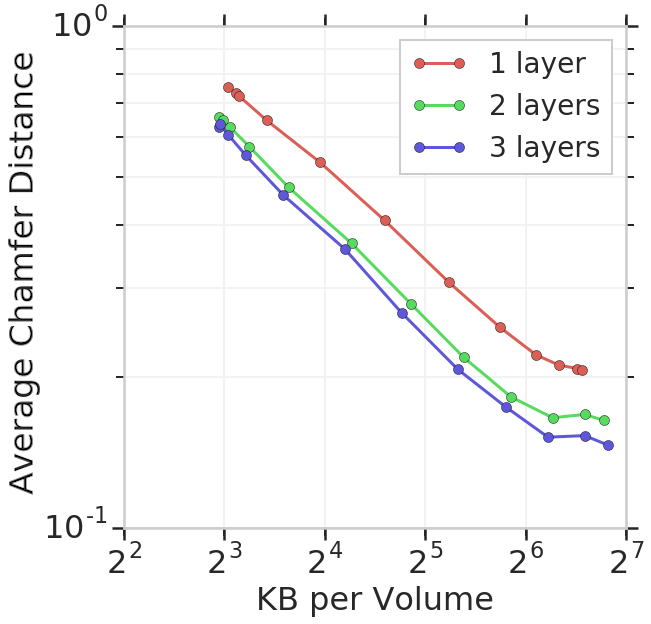}
\end{center}
\vspace{-1.5em}
\caption{Number of layers.}
\end{subfigure}
\begin{subfigure}{0.49\linewidth}
\begin{center}
    \includegraphics[width=1.0\linewidth]{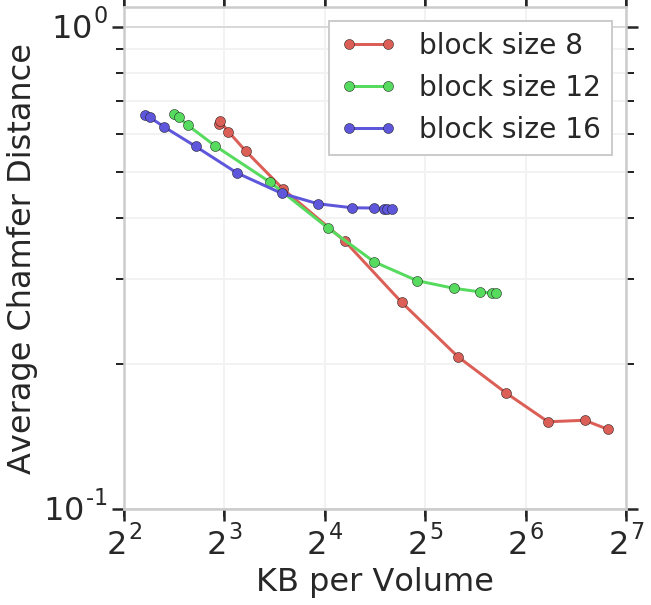}
\end{center}
\vspace{-1.5em}
\caption{Different block sizes.}
\end{subfigure}
\vspace{-.5em}
\caption{\textbf{Ablation studies:} (a) Larger number of layers in both the encoder and the decoder improves performance, although with diminishing returns and increasing model size. (b) Larger block size performs better at low rates, while smaller blocks achieve better trade-off at higher rates.
}
\vspace{-1em}
\label{fig:exp_ablation}
\end{figure}

\paragraph{Ablation studies}
The impact of network architecture on compression is evaluated in \Fig{exp_ablation}.
While having more layers leads to better results, there are diminishing returns. To keep the model size practical, we restricted our model to three layers (${<}1.8$MB).
We also perform ablation for the block-size (voxels/block). Since in all volumes, the voxel size is $5$~mm, a block with block-size $8^3$ has the physical size of $40\text{mm}^3$.
Note that increasing the size of each block reduces the number of blocks.
Results show that if one has a budget of more than $12$~KB per volume, using block size $8^3$ yields much better rate-distortion performance.
Therefore in the following experiments, $\times 3$ layers with $8^3$ blocks is used.

\begin{figure}
\begin{subfigure}{0.49\linewidth}
\begin{center}
    \includegraphics[width=1.0\linewidth]{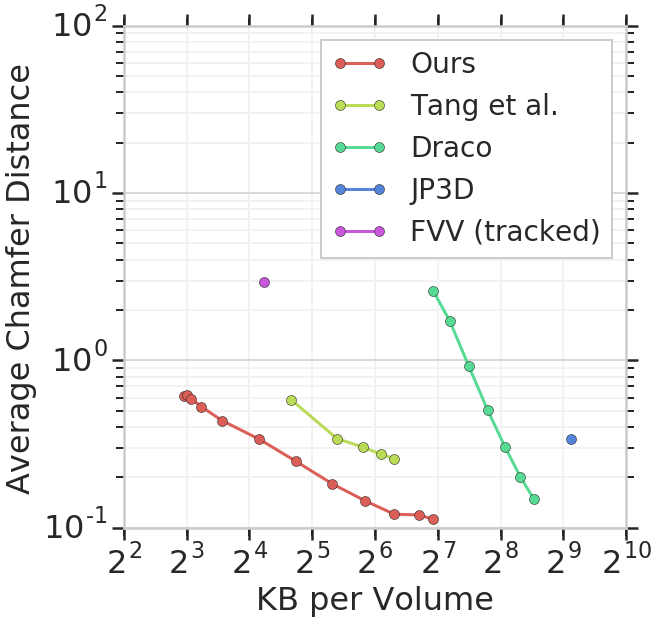}
    \label{fig:exp_beaming}
    \vspace{-1.5em}
    \caption{Dataset~\citet{compression}}
\end{center}
\end{subfigure}
\begin{subfigure}{0.49\linewidth}
\begin{center}
    \includegraphics[width=1.0\linewidth]{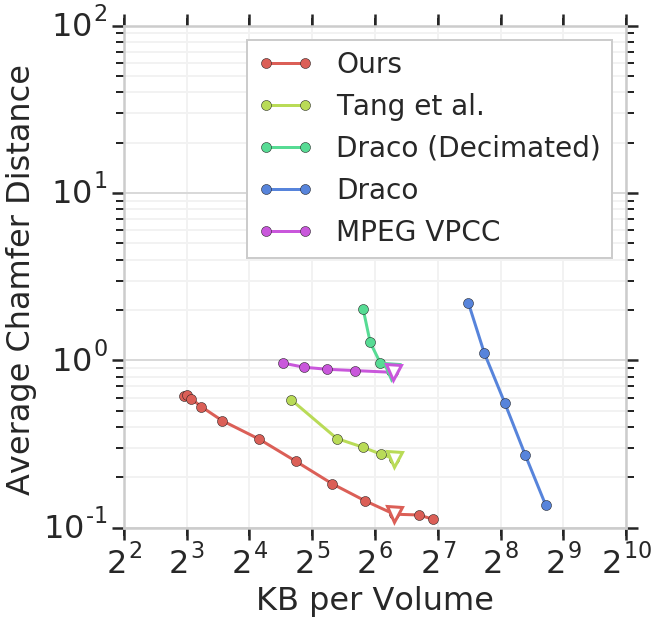}
    \label{fig:exp_holodeck}
    \vspace{-1.5em}
    \caption{Dataset~\citet{relightables}}
\end{center}
\end{subfigure}
\vspace{-.5em}
\caption{
\textbf{Quantitative comparisons --}
Our method yields the best rate-distortion among state-of-the-arts. Data points marked with $\triangledown$ are selected to have similar rates and whose distortion is visualized qualitatively in~\Figure{exp_qualitative}.
}
\label{fig:exp_arts}
\vspace{-1em}
\end{figure}

\paragraph{State-of-the-art comparisons}
We compare with state-of-the-art geometry compression methods, including
two volumetric methods: \citet{compression} and JP3D~\cite{jpeg2000_jp3d}; 
two mesh compression: Draco~\cite{draco} and Free Viewpoint Video (FVV)~\cite{fvv}; 
as well as a point cloud compressor MPEG VPCC~\cite{schwarz2018emerging}.
See their parameters in the \supplementary{}.
For most of the methods, we sweep the rate hyper parameter to generate rate-distortion curves.
The dataset~\cite{relightables} contains high-resolution meshes~(${\sim}250$K vertices), which has a negative impact on the Draco compression rate.
Hence, for Draco only, we decimate the meshes to 25K vertices termed as \textit{Draco~(decimated)} to make it comparable to other methods.
\Fig{exp_arts} shows that on both datasets, our method significantly outperforms all prior art in both rate and distortion.
For instance, to achieve the same level of rate (marked with $\triangledown$ in \Fig{exp_arts} (b)), the distortion of our method~($0.12$) is $50\%$ of \citet{compression}~($0.25$), and $14\%$ of Draco (decimated) ($0.86$) and MPEG ($0.84)$. To achieve the same distortion level ($0.25$), our method~($26$KB) only requires $33\%$ of the previous best performing method~\citet{compression}~($79$KB).

To showcase difference in distortion, we select a few qualitative examples with similar rates, and visualize them in \Figure{exp_qualitative}: the Draco (decimated) results are low-res, the MPEG V-PCC results are noisy, while the results of~\citet{compression} suffer blocking artifacts.

\paragraph{Efficiency} To assess the complexity of our neural network, we measure the runtime of the encoder and the decoder. We freeze our graph and run it using the Tensorflow C++ interface on a single NVIDIA PASCAL TITAN Xp GPU. Our range encoder implementation is single-threaded CPU code, hence we include only the neural network inference time.
We measure 20~ms to run \textit{both} encoder and decoder on all the blocks of a single volume.

\begin{figure}
\begin{center}
    \includegraphics[trim={0cm 3.5cm 7cm 0cm},clip,width=1.0\linewidth]{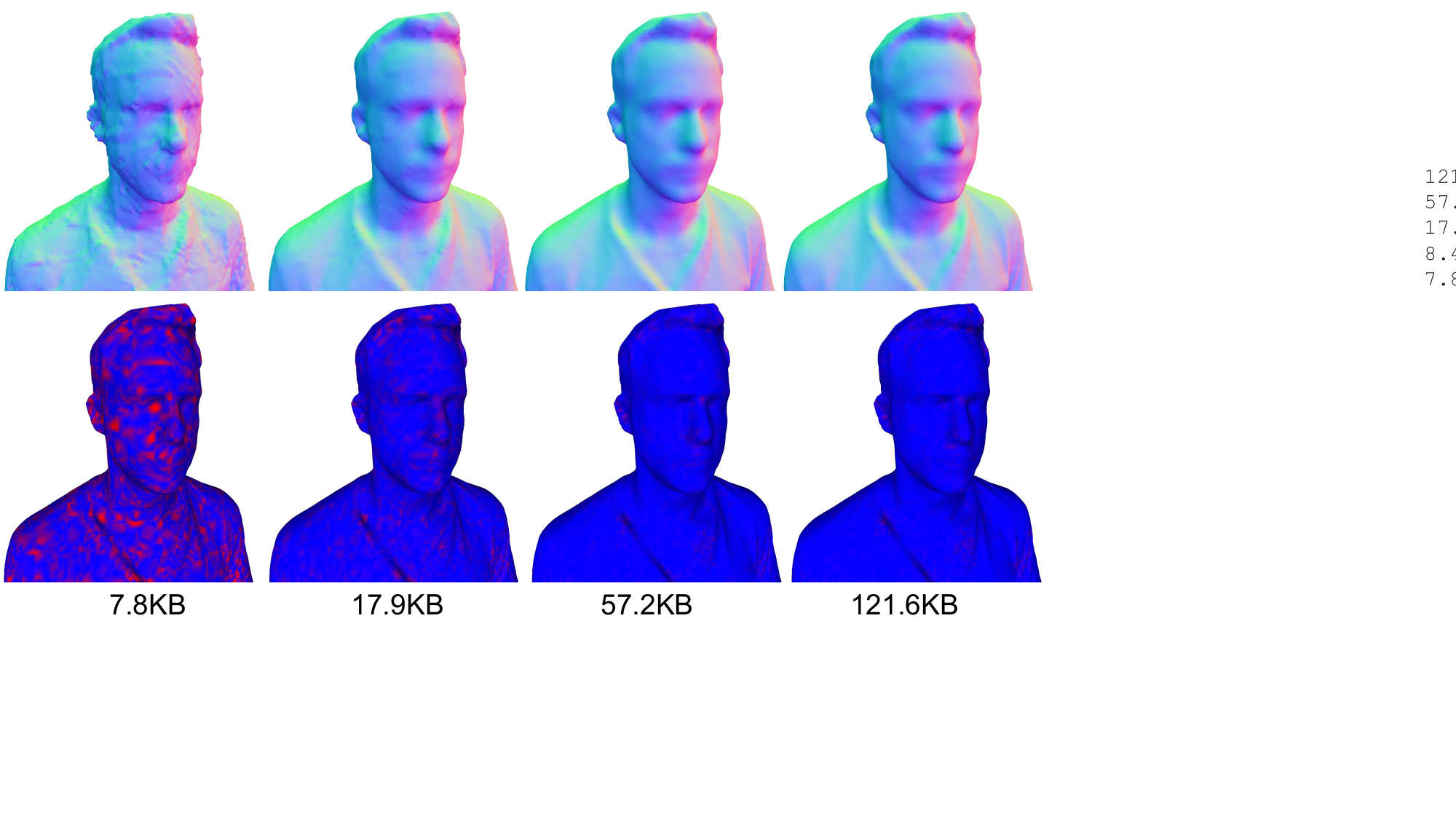}
    \vspace{-3em}
\end{center}
\caption{\textbf{Geometry / Qualitative -- } Examples from the~\citet{relightables} dataset with different rates. (1st row) Decompressed meshes. (2nd row) Shortest distance from decompressed vertices to ground truth surface. Distance between $[0, 2.5\text{mm}]$ is mapped to $[0, 255]$ on the red channel.}
\label{fig:exp_rate_qualitative}
\vspace{-1em}
\end{figure}

\subsection{Texture compression}
\label{sec:texture_experiments}
\begin{figure}
\includegraphics[trim={0cm 2.5cm 3cm 0cm},clip, width=1.0\linewidth]{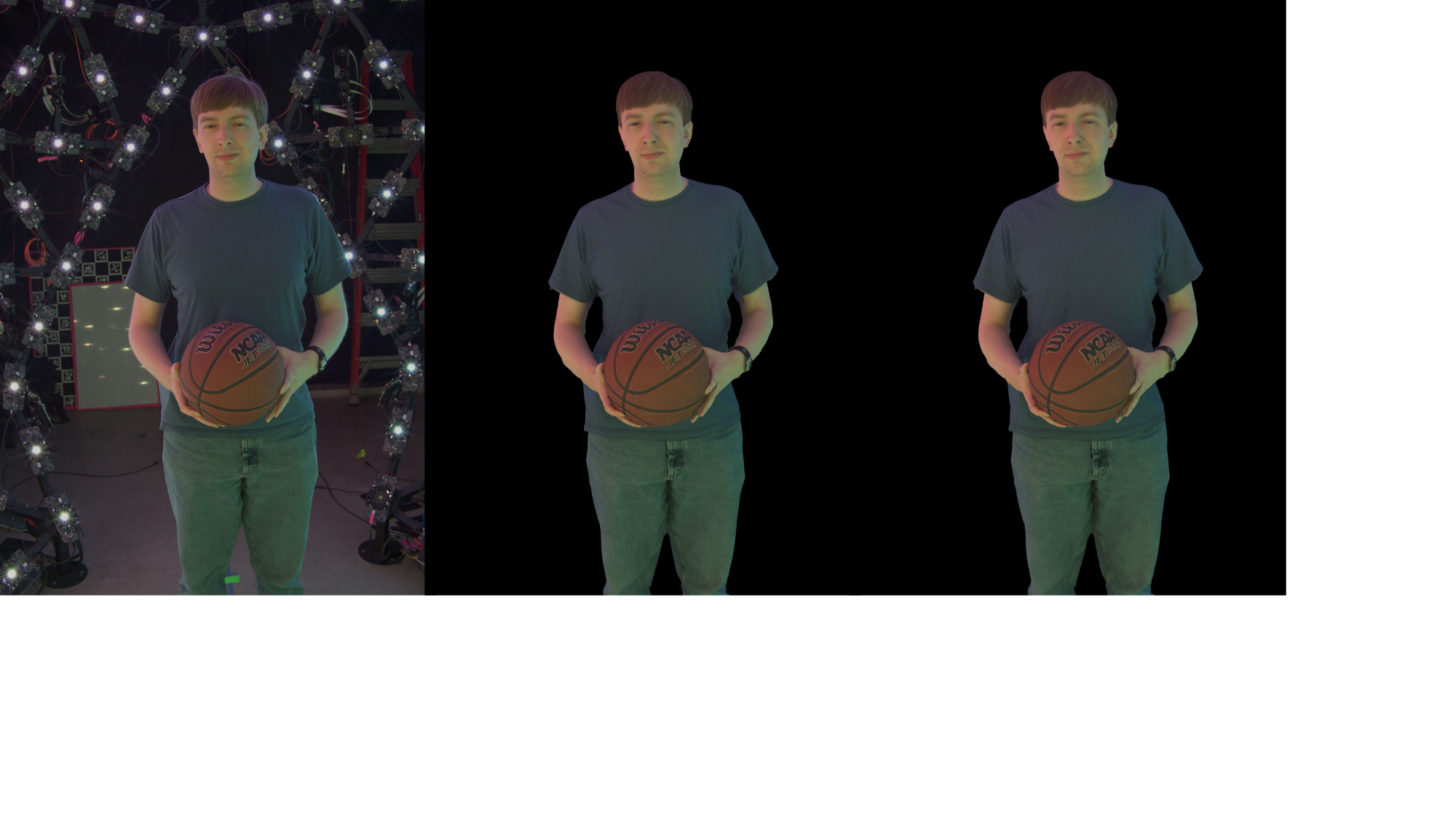}
\vspace{-3em}
\caption{
\textbf{Texture / Qualitative -- }
A frame taken from the comparison sequences in the \video{}: (left) raw rgb image from camera; (mid) rendered with UVAtlas~\cite{zhou2004iso}; (right) rendered with our texture atlas.
there is no visible difference in quality.
}
\label{fig:exp_texture_qualitative}
\vspace{-1.5em}
\end{figure}

\begin{figure*}[t!]
\begin{center}
\includegraphics[trim={0cm 1cm 8cm 0cm},clip,width=1.0\linewidth]{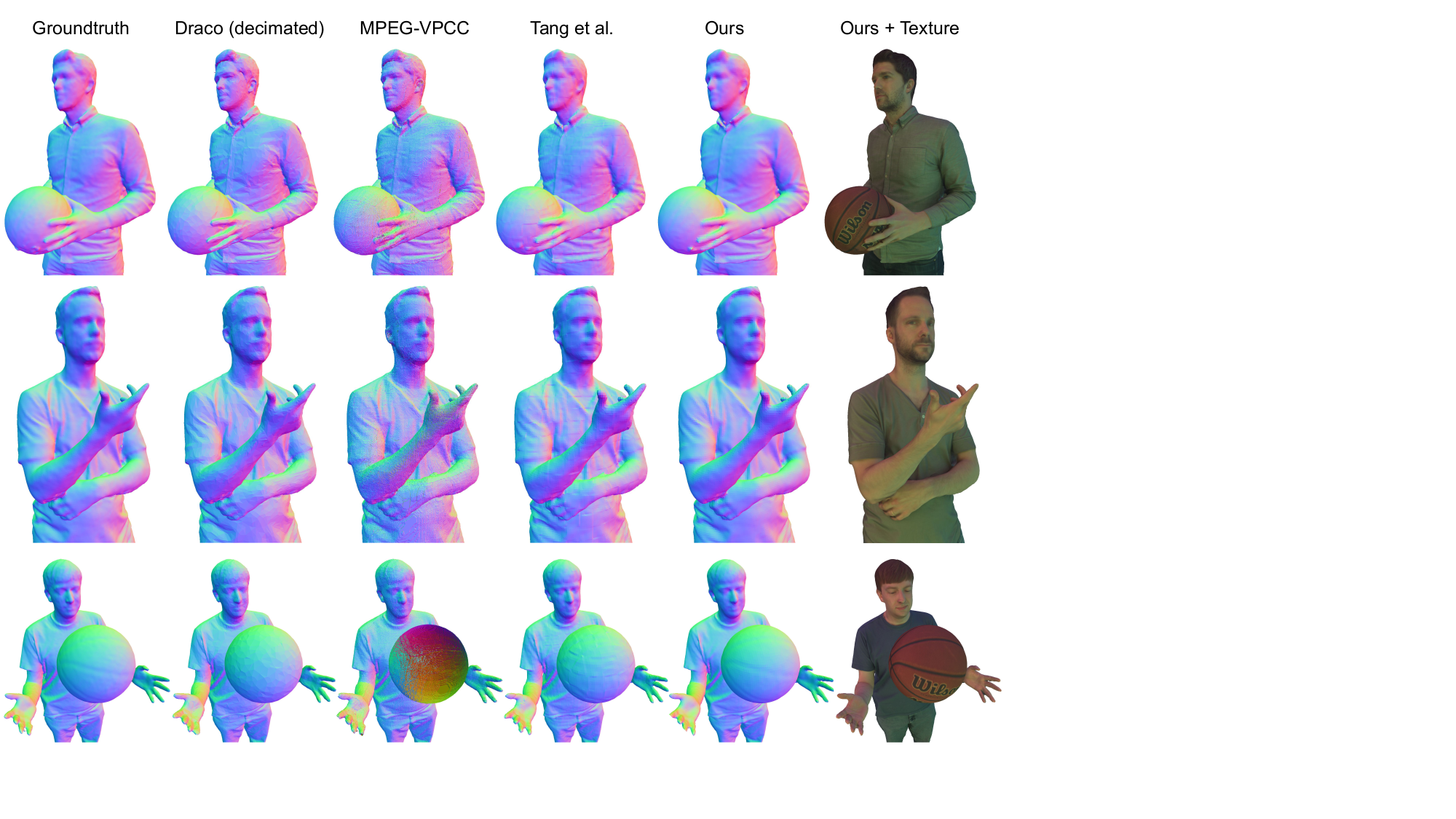}
\vspace{-1em}
\caption{\textbf{Qualitative vs. State-of-the-art -- }
Examples are selected to have a similar rate but different distortions, which correspond to the markers in \Fig{exp_arts}~(right) -- \textit{flat Phong shading} is used in all cases to reveal artifacts. In order to achieve the same level of bitrate as other methods, Draco requires decimating input, which results in low-poly reconstruction. MPEG-VPCC only compresses point clouds. \citet{compression} has visible block artifact. Our method achieves the best distortion. }
\label{fig:exp_qualitative}
\vspace{-1em}
\end{center}
\end{figure*}

We compare our texture parametrization to UVAtlas~\cite{zhou2004iso}. In order to showcase the benefit of Morton packing, we also have a block-based baseline where na\"ive bin packing is used without any spatio-temporal coherence, as shown in \Table{texture_quantitative}.
To preserve the high quality of the target dataset ~\cite{relightables}, we generate high-res texture maps~(4096x4096) for all experiments.
The texture maps of each sequence are compressed with the \texttt{H.264} implementation from \texttt{FFMpeg} with default parameters.
Per-frame compressed sizes of different methods are reported to showcase how texture parametrization impacts the compression rate.
In order to measure distortion, each textured volume with its decompressed texture atlas is rendered into the viewpoints of RGB cameras that were used to construct the volumes, and compared with the corresponding raw RGB image.
For simplicity we only select 10 views (out of 58) where the subject face is visible.
When computing distortion, masks are used to ensure only foreground pixels are considered, as shown in \Fig{exp_texture_qualitative}.

\begin{table}[ht]
\centering
\begin{tabularx}{0.475\textwidth}{Xrrrr}
\toprule
\textbf{Method} & \textbf{Rate} & \textbf{PSNR} & \textbf{SSIM} & \textbf{MS-SSIM} \\
\midrule
UVAtlas~\cite{zhou2004iso} & 457 & 30.9 & {0.923} & {0.939}  \\
Ours (Na\"ive) & 529 & 30.9 & 0.924 & {0.939} \\
Ours (Morton) & \textbf{350} & 30.9 & 0.924 & 0.940\\
\bottomrule   
\end{tabularx}
\vspace{-.5em}
\caption{
\textbf{Texture / Quantitative} -- Average  KB per volume from video compression is reported as Rate. With negligible difference in distortion under different metrics (PSNR, SSIM~\cite{wang2004image} and MS-SSIM~\cite{wang2003multiscale}), our method preserves better spatio-temporal coherence and thus has better compression rate. See qualitative results in the \video{}.
}
\label{tab:texture_quantitative}
\vspace{-1.5em}
\end{table}

\section{Conclusions}
We have introduced a novel system for the compression of TSDFs and their associated textures achieving state-of-the-art results.
For geometry, we use a block-based learned encoder-decoder architecture that is particularly well suited for the uniform 3D grids typically used to store TSDFs. To train better, we present a new distortion term to emphasize the loss near the surface. Moreover, ground truth signs of the TSDF are losslessly compressed with our learned model to provide an  error bound during decompression. For texture, we propose a novel block-based texture parametrization algorithm which encourages spatio-temporal coherence without tracking and the necessity of UV coordinate compression.
As a result, our method yields a much better rate-distortion trade-off than prior art, achieving $50\%$ distortion, or when distortion is fixed, $33\%$ bitrate of~\citet{compression}.

\paragraph{Future work}
There are a number of interesting avenues for future work.
In our architecture, we have assumed blocks to be i.i.d., and dropping this assumption could further increase the compression rate~{--}~for example, one could devise an encoder that is particularly well suited to compress ``human shaped'' geometry.
Further, we do not make any use of temporal consistency in 4D sequences, while from the realm of video compression we know coding \textit{inter}-frame knowledge provides a very significant boost to compression performance.
Finally, while our per-block texture parametrization is effective, it is not included in our end-to-end training pipeline~{--}~one could learn a per-block parametrization function to minimize screen-space artifacts.

\clearpage
{
    \small
    \newpage
    \setlength{\bibsep}{0pt}
    \bibliographystyle{plainnat}
    \bibliography{biblio}
}
\clearpage
\title{Deep Implicit Volume Compression \\ (Supplementary Material)}

\author{Danhang Tang\footnotemark[1] \quad Saurabh Singh\footnotemark[1]
\quad Philip A.~Chou \quad Christian H{\"a}ne \quad 
Mingsong Dou \\ Sean Fanello \quad Jonathan Taylor \quad Philip Davidson \quad Onur G.~Guleryuz
\quad Yinda Zhang \\ Shahram Izadi \quad Andrea Tagliasacchi \quad Sofien Bouaziz \quad Cem Keskin
\\[.5em]
Google
}

\maketitle

\section{Background on compression}
\label{sec:background}

\paragraph{Truncated Signed Distance Fields}
A surface $\surface$ represented in TSDF implicit form is the zero crossing of a function $\tsdf(\block){:} \R^3 {\rightarrow} \R$ that interpolates a uniform $\volumeDomain{}$ 3D grid of truncated (and signed) distances from the surface.
By convention, distances outside and inside the surface get positive and negative signs respectively, and magnitudes are truncated by a threshold value $\threshold$.
Typically a method like marching cubes~\cite{marchingcubes} is used to determine the \textit{topology} of each voxel (\ie which voxel edges intersect with the surface), as well as the offsets of the intersection points for the valid edges, which are then used to form a triangular mesh.

\paragraph{Lossless compression}
The primary goal of general purpose lossless compression is to minimize the storage or transmission costs (typically measured in bits) of a discrete dataset $\mathcal{X}=(\block_1,\ldots,\block_N)$.
Each data point of $\mathcal{X}$ is mapped to a variable length string of bits for storage or transmission by the sender. 
A receiver then inverts the mapping to recover the original data from the transmitted bits. 
The Shannon entropy $H{=}-\sum_{\block}p_{\block}(\block)\log p_{\block}(\block)$ provides an achievable lower bound on the rate, \ie the minimum expected number of bits required to encode an element, where $p_{\block}(\block)$ is the underlying distribution of $\block$.
This is achievable by encoding $\block$ to a bit string of length $-\log p_{\block}(\block)$ bits.
Although this length is not necessarily an integer, it can be achieved arbitrarily closely on average by an arithmetic coder~\cite{cover2006}.  With this encoding, the number of bits needed to code the entire dataset is
\begin{align}
    R(\mathcal{X}) = -\tfrac{1}{N} \sum_{i=1}^N \log p_{\block}(\block_i),
\end{align}
where $R$ is referred to as the \emph{bit rate} of the compression. 

\paragraph{Lossy compression}
In contrast, lossy compression methods can achieve significantly higher compression rates by allowing errors in the received data.
These errors are typically referred to as \emph{distortion} $D$. %
In lossy compression there is a fundamental compromise between the distortion $D$ and the bit rate $R$, referred to as \textit{rate-distortion} trade--off, where distortion can be decreased by spending more bits.
Minimizing $D$ subject to a constraint on $R$ leads to the following unconstrained optimization problem \cite{ChouLG:89a, OrtegaR:98}
\begin{align}
\label{eq:org_rate_dist}
\argmin_{\hat{\block}} D(\block, \hat{\block}) + \lambda R(\hat{\block}),
\end{align}
where $\hat{\block}$ is a discrete lossy representation of $\block$ and $\lambda$ is a trade--off parameter. 
Higher values of $\lambda$ result in better bit rates at the expense of increased distortion.

\paragraph{Lossy transform coding}
Often $\block$ is high dimensional, making the direct optimization of the problem above intractable.
As a result, \emph{lossy transform coding} is more commonly used instead. 
In lossy transform coding, 
a transformation is used to transform the original data $\block$ into a latent representation $\code{=}\encoder(\block; \ptheta_e)$ and another is used to approximately recover the original data $\hat{\block}{=}\decoder(\quantizedcode; \ptheta_d)$ from the lossy latent representation $\quantizedcode$.
The transformations $\encoder$ and $\decoder$, with parameters $\ptheta_e$ and $ \ptheta_d$, respectively, are typically chosen to simplify the conversion from $\code$ to its lossy \textit{discrete} version $\quantizedcode{=}Q(\code)$~{--}~a process called \emph{quantization}.
While $\encoder$ and $\decoder$ can be invertible transformations (\eg the discrete cosine transform used for JPEG compression), in general they are not required to be.
Thus, with $\ptheta{=}\{\ptheta_e, \ptheta_d\}$, the original rate-distortion problem can be re-written as
\begin{align}
\label{eq:rate_dist}
\argmin_{\ptheta, \pphi} \:\:
D(\block, \hat{\block}; \ptheta) + 
\lambda R(\hat{\code}; \pphi),
\end{align}
where $\hat{\block}{=}\decoder(\quantizedcode; \ptheta_d)$, $\quantizedcode{=}Q(\encoder(\block; \ptheta_e))$, and the bit rate is $R(\quantizedcode; \pphi){=}{-}
\log p_{\quantizedcode}(\quantizedcode; \pphi)$, with $p_{\quantizedcode}$ as a probability model of~$\quantizedcode$ with parameters~$\pphi$ that is learned jointly with~$\ptheta$. 
The code $\quantizedcode$ is converted to the corresponding variable length bit representation by entropy coding using the learned prior distribution $p_{\quantizedcode}$.

\begin{table*}[h!]
\centering
\begin{tabularx}{\textwidth}{|l|l|X|}
\toprule
\textbf{Method} & \textbf{Rate Parameters (varied)} & \textbf{Fixed Parameters} \\
\midrule
Ours & $\lambda=\frac{1}{10^{\mu}}, \mu=i\times\frac{\log_{10}200000}{11}$ (for $ i=0,\dots11$) &   \\
\midrule
\citet{compression} & $K_{total} = 1024, 2048\ldots5120$ & $\text{numRetainedKLTBases}=64$\\
\midrule
Google Draco~\cite{draco} & $\text{qp}=8,\ldots,11$ &
$\text{qt}=11$\newline $\text{skip}=\text{normal}$  \\
\midrule
MPEG V-PCC~\cite{schwarz2018emerging} & r$i$ configurations (for $i=1,\ldots,5$) & $\text{geometry3dCoordinatesBitdepth}=11$ \newline 
$\text{geometryNominal2dBitdepth}=8$ \newline $\text{minNormSumOfInvDist4MPSelection}=0.36$ \newline
$\text{partialAdditionalProjectionPlane}=0.15$ \newline 
$\text{minimumImageWidth}=2560$ \newline
$\text{apply3dMotionCompensation}=0$ \\
\bottomrule   
\end{tabularx}
\vspace{-.5em}
\caption{Parameters used for the experiments in Figure 7 of the main paper.}
\label{tab:parameters}
\end{table*}

\paragraph{Quantization} 
Since the quantization operation is non-differentiable, training such a network in an end-to-end fashion is challenging.
\citet{balle2017end} propose simulating quantization noise during training rather than explicitly discretizing the code.
Specifically, they quantize $\code$ by rounding to nearest integer $\quantizedcode{=}Q(\encoder(\block; \ptheta_e)){=}\lfloor \encoder(\block; \ptheta_e) \rceil$, which they model by adding of uniform noise during training, \ie $\quantizedcode{=}\encoder(\block; \ptheta_e){+}\epsilon$, $\epsilon{~}\sim~\mathcal{U}[-0.5, 0.5]$ to simulate quantization errors; see~\cite{balle2017end} for additional details.

\begin{figure*}[t]
\begin{center}
\includegraphics[trim={0cm 0cm 0cm 0cm},clip,width=\linewidth]{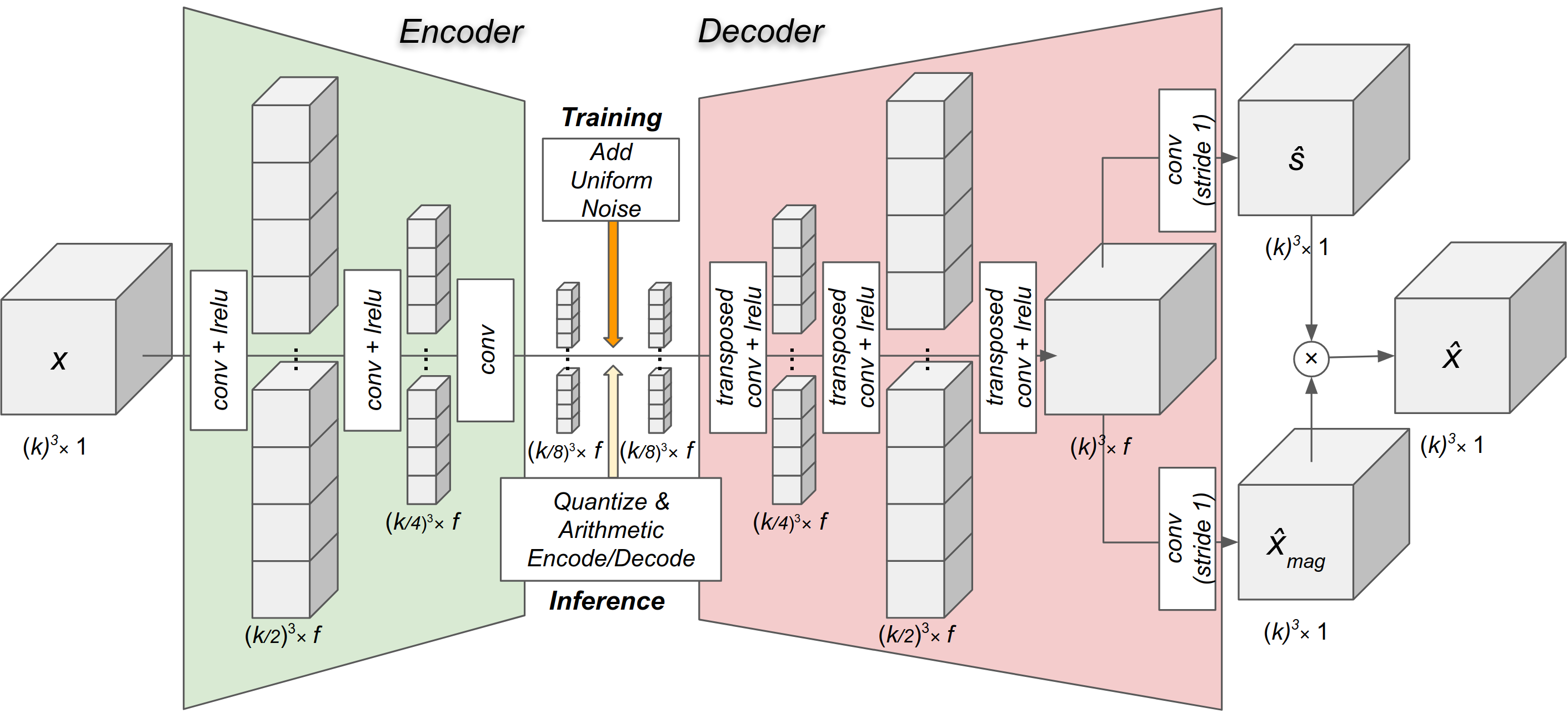}
\vspace{-1em}
\caption{\textbf{Network architecture.} The encoder $\encoder$ consists of three convolutional layers and the decoder $\decoder$ has three transposed convolution layers, each with a stride of two. $\decoder$ has two convolutional heads with a stride of one, which separates sign prediction from TSDF estimation. Refer to Section 4 in main paper for details.}
\label{fig:network_architecture}
\vspace{-1em}
\end{center}
\end{figure*}

\begin{figure*}[t]
\begin{center}
\includegraphics[trim={0cm 0cm 0cm 0cm},clip,width=0.65\linewidth]{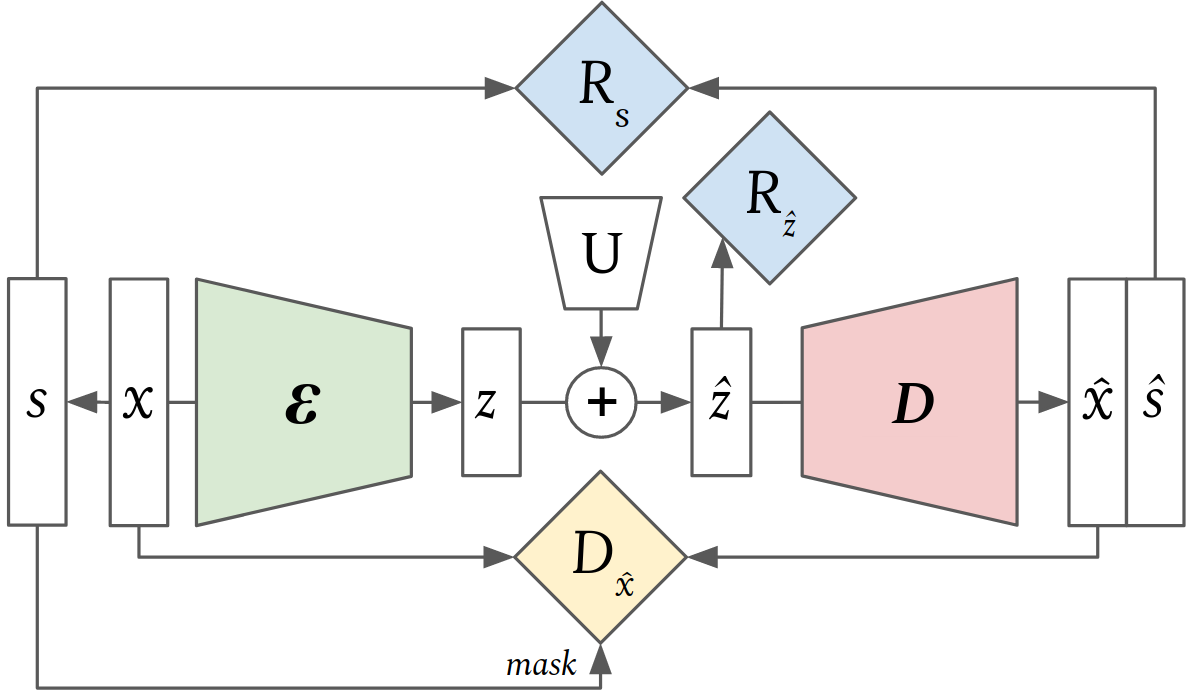}
\vspace{-1em}
\caption{\textbf{Training losses.} During training, we employ three different losses as explained in~\Section{geometry}. Here, the distortion loss $\lossdist$ makes use of the ground truth signs $\sign$ to mask the voxels that have no neighboring voxels with opposing signs and have therefore less significance. $\losssign$ is the cross entropy between the predicted and actual signs, which is used to minimize the bit rate for compressed ground truth signals. $\lossrate$ is an estimate of the differential entropy of the noisy latent code, also used to minimize the bit rate for the compressed latent code $\quantizedcode$.}
\label{fig:training_losses}
\vspace{-1em}
\end{center}
\end{figure*}

\section{Network architecture and training}
We visualize the architecture of our model in ~\Figure{network_architecture}, which is formed by a three layer encoder and decoder. While the architecture is similar to a convolutional autoencoder (implemented with convolutions in the encoder and transposed convolutions in the decoder), the main difference lies in the transformation the latent code goes through, and the additional losses that aim to minimize the bit rate as well as the reconstruction error, as visualized in ~\Figure{training_losses}. Specifically, we add uniform noise to the code during training to simulate quantization. At test time we quantize the code and compress it with an entropy coder. Additionally, the decoder has two final convolutional heads that separate the estimation of signs and the TSDF values. The one and two layer models we experiment with are similar with fewer layers. 

Figure~\ref{fig:training_losses} provides an overview of our training setup with the dependencies for the three terms in our training loss. Unlike a regular autoencoder which only aims to minimize the reconstruction error, we employ two additional losses $\lossrate$ and $\losssign$ to minimize the bit rates for the compressed signals for the latent code  and the ground truth signs. Additionally, instead of equally weighting each element the reconstructed $\hat{\block}$, we use the ground truth signs $\sign$ to mask the voxels that have no neighboring voxels with opposing signs and have therefore less significance. 

\section{Baseline parameters}
\label{sec:parameters}

The parameters used in our experiments (Figure 6) are described in~\Table{parameters}, except for JP3D~\cite{jpeg2000_jp3d} and FVV~\cite{fvv} which we obtained from~\citet{compression}. To generate a curve, we varied the corresponding rate parameter during inference, whilst keeping other parameters fixed as shown. Notations and definitions of parameters can be found in respective citations.

\end{document}